\documentclass[a4paper,oneside,article]{memoir}

\setlrmarginsandblock{1.5in}{1.5in}{*}
\setulmarginsandblock{1.5in}{1.5in}{*}
\checkandfixthelayout

\usepackage[utf8]{inputenc}
\usepackage[T1]{fontenc}

\usepackage{siunitx}
\DeclareSIUnit\fm{fm}
\DeclareSIUnit\feV{feV}
\DeclareSIUnit\aeV{aeV}

\usepackage{dblfloatfix}
\usepackage{graphicx}
\graphicspath{{figures/}}
\usepackage{mhchem}
\usepackage{isotope}
\usepackage[draft]{fixme}
\usepackage[pdftex]{color}
\usepackage[font=footnotesize,labelfont=bf]{caption}
\usepackage[font=footnotesize,labelfont=bf]{subcaption}
\usepackage{lineno,hyperref}
\usepackage{cleveref}
\usepackage{booktabs}

\newcommand{\Li}{\isotope[7]{Li}}
\newcommand{\Be}{\isotope[8]{Be}}
\newcommand{\Boron}{\isotope[8]{B}}

\begin{document}

\title{R-Matrix parametrization for $\gamma$ decays to unbound states}

\author{Michael Munch\\Institute of Physics and Astronomy, Aarhus University, Denmark}
\maketitle

\chapter{Introduction}
R-matrix theory was originally developed to describe nuclear reactions \cite{Lane1958}.
The framework was further extended to describe $\beta$ decay to unbound states \cite{Barker1969,Warburton1986}. However, at the
time of writing, no clear description of $\gamma$ decays to unbound states exist. Such a description
will be presented in this note. 

\chapter{R-matrix parameterization}
\label{sec:r-matr-param}

The motivation for this note is the following reaction
\begin{equation}
  \label{eq:reaction}
  p + \Li \rightarrow \Be^{*} \rightarrow \gamma + \Be^{**} \rightarrow \gamma + \alpha + \alpha.
\end{equation}
It will be assumed that the initial reaction
proceeds through a single isolated resonance in the compound system. Under these conditions,
the treatment is quite similar to that of $\beta$
decays of \Boron{}
and \isotope[8]{Li} \cite{Barker1969,Warburton1986}. The notation is that of
Ref.\ \cite{Lane1958} unless stated otherwise.

\begin{figure}[h]
  \centering
  \includegraphics[width=0.75\columnwidth]{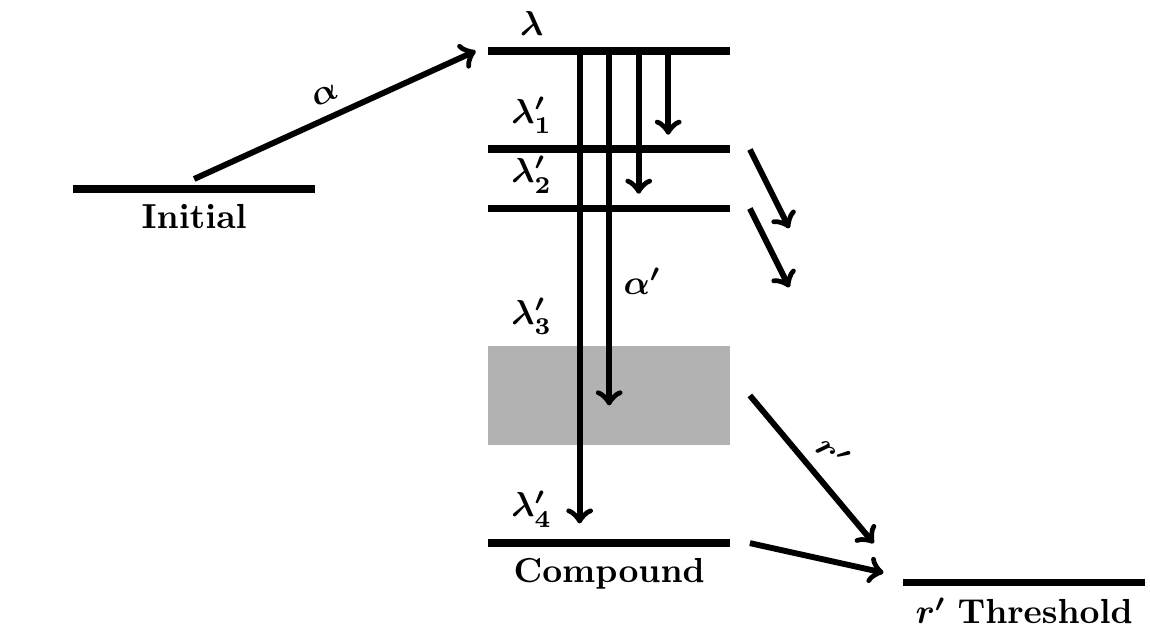}
  \caption{Sketch of the general decay scheme. A resonance $\lambda$ in the compound system is formed
  via channel $\alpha$. This resonance decays to another unbound compound level $\lambda'$ via a $\gamma$-channel
  $\alpha'$. The unbound level subsequently decays via $r'$.
  }
  \label{fig:rmat-scheme}
\end{figure}

Using the formalism in Ref.\ \cite[XIII.2]{Lane1958} the reaction is described as a sequential
process. \Cref{fig:rmat-scheme} illustrates the general case in which
an isolated initial level $\lambda$
is populated via channel%
\footnote{Here $\alpha$ and $\alpha'$ is the notation used for a channel not an $\alpha$ particle.}
$\alpha$
and decays via channel $\alpha'$
to an unbound fragment in a state $\lambda'$
and internal energy $E_{2}'$,
which finally decays via the open channel $r'$. The differential cross section for this
reaction is given as
\begin{equation}
  \label{eq:cross1}
  \frac{d\sigma_{\alpha \alpha'}(E_{2}'r')}{dE_{2}'} = \frac{\pi}{k_{\alpha}^{2}} \sum_{c\,c'} 
  \frac{g_{J}\ \Gamma_{\lambda c}\,\delta\Gamma_{\lambda' c'}(E_{2}'r')}{(E_{\lambda} + \Delta_{\lambda} - E)^{2} + (\Gamma_{\lambda}/2)^{2}},
\end{equation}
where $c$
(and $c'$)
specifies the two subsystems, their spins, the channel spin $s$
and their relative angular momentum $\ell$.
Note that the notation in Ref.\ \cite{Lane1958} is ambiguous, so here we use
$\delta\Gamma_{\lambda' c'}(E_{2}'r') \equiv \frac{d\Gamma_{\lambda'c'}(E_{2}'r')}{dE_{2}'}$.

Generally, the integral of $\delta\Gamma_{\lambda' c'}(E_{2}'r')$
should be included in $\Gamma_{\lambda}$,
however, the photon channels do not contribute significantly, so the width of the initial
level can be approximated with the particle width
$\Gamma_{\lambda} \approx \sum_{c_{p}} \Gamma_{\lambda c_{p}}$. Additionally, for a narrow entry
level, this can be rewritten into the familiar Breit-Wigner form
\begin{equation}
  \label{eq:cross2}
  \frac{d\sigma_{\alpha \alpha'}(E_{2}'r')}{dE_{2}'} = \frac{\pi}{k_{a}^{2}} \sum_{s\ell s'\ell'} g_{J}
  \frac{\Gamma_{\lambda c}^{0}\,\delta\Gamma_{\lambda' c'}^{0}(E_{2}'r')}{(E_{\lambda}^{0} - E)^{2} + (\sum_{c_{p}} \Gamma_{c_{p}}^{0}/2)^{2}},
\end{equation}
where parameters with superscript 0 are so-called observed parameters \cite{Brune2002}.

Due to their small coupling to the nucleus, photon channels can be included in R-matrix theory
using a perturbation approach. Ref.\ \cite[XIII.3]{Lane1958} has derived the theory for one
photon without damping, which applies to this case.  Within this framework particle and photon
channels can be treated almost identically, except photon channels should not be included in the
level matrix. Additionally, one can work directly with observable parameters if one uses
the 
Brune formalism \cite{Brune2002}. With this formalism the level matrix is defined as
\begin{align}
  \label{eq:brune}
  [\tilde{\mathbf{A}}^{-1}]_{\lambda \mu} =& (\tilde{E}_{\lambda} - E) \delta_{\lambda\mu} \notag
                                     - \sum_{c} \tilde{\gamma}_{\lambda c}\tilde{\gamma}_{\mu c} (S_{c} + iP_{c}) \\
  +& \sum_{c}
     \begin{cases}
       \tilde{\gamma}_{\lambda c}^{2} S_{\lambda c}  &\text{for } \lambda=\mu,\\
       \tilde{\gamma}_{\lambda c} \tilde{\gamma}_{\mu c}
       \frac{S_{\lambda c} (E-\tilde{E}_\mu) - S_{\mu c}(E-\tilde{E}_\lambda)}{\tilde{E}_\lambda - \tilde{E}_\mu} &
       \text{for }\lambda \neq \mu,
     \end{cases}
\end{align}
where $\tilde{E}$
is the observable resonance energy and $\tilde{\gamma}_{\lambda c}$
the reduced width coupling of a level $\lambda$ to a channel $c$. $P_{c}$ and $S_{c}$ are respectively
the penetrability and shift function as defined in regular R-matrix theory \cite{Lane1958} with
$S_{\lambda c} = S_{c}(\tilde{E}_{\lambda})$.

The partial width for emission of a $L$-wave
photon with energy $E$
from a level $\lambda$ is related to its reduced width via $\Gamma_{\lambda L} = 2E^{2L+1}\,\gamma^{2}_{\lambda L}$. The
partial width for a particle is given as $\Gamma_{\lambda c} = 2P_{c}\,\gamma^{2}_{\lambda c}$. In order to have a
symmetric notation we define the $\gamma$ penetrability as $P_{L} \equiv E^{2L+1}$.

Following Ref.\ \cite{Barker1988} we adopt the following expression to describe the differential
partial width of the intermediary state
\begin{equation}
  \label{eq:width}
  \delta\Gamma^{0}_{\lambda c'(E_{2}'r')} = \frac{2P_{c'} 2P_{r'}}{2\pi}
  \Big|\sum_{\nu\mu} \tilde{\gamma}_{\lambda c'(\nu)} \tilde{\gamma}_{\mu r'} \tilde{A}_{\nu \mu} \Big|^{2}.
\end{equation}
This reduces to (XIII 2.10) of Ref.\ \cite{Lane1958} for the case of a single isolated
intermediary resonance,
\begin{equation}
  \label{eq:width1}
  \delta\Gamma^{0}_{\lambda c'(E_{2}'r')} = \frac{1}{2\pi}\ 
  \frac{\tilde{\Gamma}_{\lambda c'(\lambda')}\tilde{\Gamma}_{\lambda' r'}}{(\tilde{E}_{\lambda'} + \Delta_{\lambda'} - E_{2}')^{2} + (\tilde{\Gamma}_{\lambda'}/2)^{2}}.
\end{equation}

The observed partial decay width of $\lambda$
to a specific resonance $\lambda'$ is then given as the integral over the resonance peak. 
For an isolated intermediary level the shift function is approximately linear over the
resonance and the integral can be performed analytically
\begin{equation}
  \label{eq:width}
  \Gamma^{0}_{\lambda c'(\lambda')} = \int_{\lambda'} \delta\Gamma^{0}_{\lambda c'(E_{2}'r')}\ dE_{2}' \approx \frac{2P_{c'}\tilde{\gamma}_{\lambda c'(\lambda')}^{2}}{1+\Sigma_{c}\tilde{\gamma}_{\lambda'c}^{2}
    \frac{dS_{c}}{dE} \big|_{\tilde{E}_{\lambda'}}}.
\end{equation}
This is identical to the expression for the observed width in Ref.\ \cite{Brune2002}.

Inserting \cref{eq:width} into \cref{eq:cross1} or \cref{eq:cross2} yields the full
expression. Contributions from the same multipolarity is summed coherently while contributions
from different multipolarities are added incoherently. 
\chapter{Conclusion}
\label{cha:conclusion}

Combining the R-matrix theory for sequential decays with a level density inspired from $\beta$
decay studies, it is possible to describe $\gamma$ decays to unbound states.

\bibliographystyle{unsrt}
\bibliography{IFA019-article}

\end{document}